\documentclass[runningheads]{llncs}
\usepackage[T1]{fontenc}
\usepackage{graphicx}
\usepackage{amsmath}
\usepackage{amssymb}
\usepackage{amsfonts}

\usepackage{booktabs}
\usepackage{multirow}
\usepackage{array}

\usepackage{float}

\usepackage{algorithm}
\usepackage{algorithmic}

\usepackage{url}

\usepackage{caption}
\usepackage{subcaption}
\usepackage[hidelinks]{hyperref}
\begin{document}
\title{Agentic Federated Learning: Rule-Based Client and Server Agents for Adaptive Training}
\titlerunning{Agentic Federated Learning}
%
\author{
Deepthy K Bhaskar\inst{1}\orcidID{0009-0009-0304-7895} \and
Binu V P\inst{2} \and
Minimol B\inst{3}
}

\authorrunning{D.K. Bhaskar et al.}

\institute{
Department of Computer Engineering, Model Engineering College, APJ Abdul Kalam Technological University, Thiruvananthapuram 695016, Kerala, India.
\email{dmdl22jan005@mec.ac.in}
\and
Department of Computer Engineering, Model Engineering College, APJ Abdul Kalam Technological University, Thiruvananthapuram 695016, Kerala, India.
\email{ binuvp@mec.ac.in}
\and
Department of Biomedical Engineering, Model Engineering College, APJ Abdul Kalam Technological University, Thiruvananthapuram 695016, Kerala, India.
\email{mini@mec.ac.in}
}
\maketitle              

\begin{abstract}
Federated Learning (FL) enables collaborative model training across distributed clients without sharing raw data, making it suitable for privacy-sensitive applications such as healthcare, finance, and edge intelligence. However, conventional FL approaches rely on static client participation and fixed aggregation strategies, which limits their effectiveness under non-IID data distributions, heterogeneous client behavior, and noisy or unreliable updates. To overcome these issuess, this paper proposes an Agentic Federated Learning (AFL) framework that integrates lightweight rule-based autonomous agents at both client and server levels. The proposed framework introduces a Client-Side Agent (CSA) that dynamically adapts local training parameters, controls participation, and evaluates update reliability, while a Server-Side Orchestrator Agent (SSOA) performs quality-aware client selection and adaptive aggregation. Unlike traditional FL methods, AFL enables context-aware decision-making during the training process, improving adaptability and robustness in dynamic distributed environments. Extensive experiments conducted on the CIFAR-10 dataset under IID, non-IID, and noisy-client settings demonstrate that AFL consistently outperforms standard baselines including FedAvg and FedProx. Experimental results show improvements in classification accuracy, convergence speed, robustness against corrupted updates, and communication efficiency. Ablation studies further confirm the complementary contributions of CSA and SSOA, while statistical analysis validates the significance of the observed gains. The proposed AFL framework demonstrates that incorporating autonomous agentic reasoning into federated learning provides an effective and practical solution for intelligent, adaptive, and robust distributed learning systems.

\end{abstract}
\section{Introduction}

In recent years, Federated Learning has gained significant attention as a privacy-preserving distributed machine learning approach that enables collaborative training while maintaining data locality. \cite{ref1}. In FL, clients train local models using their private datasets and only share model updates with a central server. This decentralized approach provides significant privacy advantages and reduces the risks associated with direct data sharing, making FL well suited for applications involving sensitive information, such as healthcare, finance, mobile edge computing, and Internet of Things (IoT) applications \cite{ref2,ref3,ref4}.

Despite its advantages, conventional FL frameworks face several practical challenges when deployed in real-world heterogeneous environments. One of the most critical issues is statistical heterogeneity, commonly referred to as non-IID (Non-Independent and Identically Distributed) data, where client datasets vary significantly in size and class distribution \cite{ref5}. Such heterogeneity often leads to unstable convergence, biased global models, and degraded learning performance. In addition, federated systems are affected by unreliable client participation, noisy updates, communication constraints, and varying computational capabilities across devices \cite{ref6}. Traditional FL algorithms such as Federated Averaging rely on fixed local training procedures and static aggregation strategies, limiting their adaptability under dynamic training conditions. 

Recent advances in adaptive and intelligent distributed learning have highlighted the importance of autonomous decision-making mechanisms in improving system robustness and flexibility \cite{ref7}. Agentic artificial intelligence introduces lightweight autonomous agents capable of observing environmental conditions, reasoning over local states, and selecting context-aware actions. These characteristics make agentic systems highly relevant for federated learning environments where clients and servers operate under uncertainty and heterogeneity. Although several studies have explored adaptive optimization and robust aggregation strategies in FL, most existing approaches still treat clients as passive entities and lack autonomous reasoning capabilities \cite{ref8,ref9}.

To address these limitations, this paper proposes an AFL framework that integrates lightweight rule-based autonomous agents at both client and server levels. The proposed framework introduces a Client-Side Agent that dynamically adapts local training hyperparameters, evaluates update reliability, and controls client participation during communication rounds. In addition, a Server-Side Orchestrator Agent performs intelligent client selection, adaptive aggregation strategy selection, and quality-aware update weighting. Unlike conventional FL methods that rely on fixed optimization rules, AFL enables context-aware adaptive decision-making throughout the training process.

The proposed AFL framework is evaluated on the CIFAR-10 benchmark dataset under IID, non-IID, and noisy-client settings to simulate realistic federated learning conditions. Experimental results demonstrate that AFL consistently improves classification accuracy, convergence speed, robustness against noisy updates, and communication efficiency compared with established FL baselines including FedAvg and FedProx. Ablation studies further validate the complementary contributions of the client-side and server-side autonomous agents.

The key contributions of this study are summarized as follows:

\begin{itemize}
    \item We propose a novel Agentic Federated Learning framework that incorporates lightweight autonomous agents into both client and server components for adaptive federated optimization.
    
    \item We design a Client-Side Agent capable of dynamic hyperparameter adaptation, participation control, and update quality assessment.
    
    \item We develop a Server-Side Orchestrator Agent that performs intelligent client selection, adaptive aggregation strategy selection, and quality-aware model update weighting.
    
    \item We conduct extensive experiments on CIFAR-10 under IID, non-IID, and noisy-client scenarios, demonstrating significant improvements in accuracy, robustness, convergence speed, and communication efficiency.
    
    \item We provide an interpretable and computationally lightweight agentic framework suitable for practical federated learning deployments in heterogeneous distributed environments.
\end{itemize}

The remainder of this paper is organized as follows. Section 2 reviews related work in federated learning, robust aggregation, and agentic intelligent systems. Section 3 presents the proposed AFL methodology. Section 4 describes the experimental setup, while Section 5 discusses the empirical results and analysis. Section 6 outlines limitations and future research directions. Finally, Section 7 concludes the paper.

\section{Related Work}

This section reviews the major research contributions related to federated learning, robust aggregation, adaptive client participation, and autonomous intelligent systems that are relevant to the proposed AFL framework. The discussion highlights the limitations of existing approaches and motivates the integration of lightweight agentic reasoning into federated learning environments.

\subsection{Federated Learning Paradigms}

FL was proposed as a privacy-preserving distributed machine learning paradigm in which participating clients collaboratively learn a shared model while retaining their data locally \cite{ref1}. By keeping sensitive data on client devices and only exchanging model parameters, FL provides strong privacy preservation and reduced communication risks. This characteristic has enabled its adoption in sensitive domains such as healthcare, finance, mobile edge computing, and Internet of Things IoT systems \cite{ref2,ref3,ref4}. Privacy preserving machine learning and secure federated optimization have also gained considerable attention in recent years. Bhaskar et al.~\cite{ref17} reviewed privacy  preserving secure machine learning approaches for distributed intelligent systems, while homomorphic encryption-based federated learning frameworks have been explored for secure medical prediction systems \cite{ref18}.

Despite these advantages, conventional FL frameworks face significant challenges under real-world heterogeneous environments. One major issue is statistical heterogeneity caused by non-IID (Non-Independent and Identically Distributed) client data distributions \cite{ref5}. Such heterogeneity often leads to unstable convergence, biased global models, and reduced learning performance. Furthermore, practical FL systems must handle unreliable client participation, communication constraints, device heterogeneity, and varying computational capabilities across distributed clients.

To address these challenges, several optimization-based approaches have been proposed. FedProx introduces a proximal regularization term to stabilize local training under heterogeneous client conditions \cite{ref10}. Other studies investigate adaptive learning rates and personalized federated optimization mechanisms to improve convergence stability. However, these methods largely rely on fixed optimization strategies and continue to treat clients as passive entities during training.

\subsection{Robust and Adaptive Aggregation}

The decentralized nature of federated learning makes it vulnerable to noisy, corrupted, or malicious client updates. Consequently, robust aggregation mechanisms have become an important research direction in FL systems. Robust aggregation methods attempt to reduce the influence of unreliable updates through statistical filtering, weighted aggregation, or Byzantine-resilient optimization \cite{ref9}.

Several aggregation techniques such as Median aggregation, Trimmed Mean, and Krum have demonstrated improved robustness against adversarial clients and noisy updates. Acar et al.~\cite{ref6} proposed debiasing mechanisms to improve personalized federated learning under heterogeneous client contributions. Recent studies have also investigated malicious client detection and secure update validation in federated environments. Bhaskar et al.~\cite{ref16} proposed a loss trend deviation detection mechanism for identifying unreliable client behavior during federated optimization. More recent approaches investigate adaptive aggregation strategies that dynamically adjust aggregation behavior according to client reliability and update consistency.

Although these approaches improve robustness, most existing methods operate using predefined aggregation protocols and static decision-making rules. They generally lack autonomous reasoning mechanisms capable of dynamically adapting participation behavior or aggregation policies during training.

\subsection{Client Selection and Participation Control}

Efficient client participation is essential for improving federated learning convergence speed, communication efficiency, and robustness. Since federated environments often involve resource-constrained and intermittently connected devices, selecting appropriate clients for each communication round is a critical challenge.

Chen et al.~\cite{ref8} introduced asynchronous federated learning strategies to support heterogeneous and resource-limited devices. Other studies have explored resource-aware client scheduling, importance-based participation, and adaptive client sampling to improve communication efficiency and training stability. Recent adaptive FL frameworks also consider client reliability and update quality during participant selection.

However, most existing approaches perform participation control exclusively at the server side and do not allow clients to autonomously determine whether participation is beneficial. As a result, unnecessary communication overhead and unstable updates may still occur under dynamic training conditions.

\subsection{Autonomous and Agentic Intelligent Systems}

Autonomous and agentic artificial intelligence systems have recently gained considerable attention due to their ability to observe environmental conditions, reason over internal states, and dynamically adapt their actions \cite{ref7}. Agentic systems support self-regulation, adaptive optimization, and context-aware decision-making, making them highly suitable for distributed intelligent environments.

Recent advances in autonomous AI have demonstrated promising applications in robotics, edge intelligence, distributed optimization, and adaptive resource management. In federated learning environments, agentic reasoning can potentially improve robustness, adaptability, and communication efficiency by enabling both clients and servers to dynamically adjust their behaviors according to evolving training conditions.

Nevertheless, the integration of autonomous agents into federated learning remains relatively underexplored. Existing FL approaches rarely incorporate lightweight agentic reasoning mechanisms capable of dynamic participation control, adaptive aggregation selection, or quality-aware orchestration. This research gap motivates the development of the proposed AFL framework.

\subsection{Recent Advances in Adaptive Federated Learning}

Recent research has increasingly focused on intelligent and adaptive federated learning strategies for heterogeneous and resource-constrained environments. Mughal et al.~\cite{ref11} proposed an adaptive federated learning framework for IoT environments using clustered edge intelligence to improve scalability and resource utilization. Cong et al.~\cite{ref12} introduced FedGA, a greedy aggregation strategy designed to improve federated learning performance under non-IID data distributions.

Recent surveys have also highlighted the growing importance of adaptive and communication-efficient federated learning for edge intelligence systems. Dritsas and Trigka~\cite{ref13} discussed recent federated learning techniques for IoT applications and emphasized the need for adaptive participation and robust aggregation strategies. Similarly, Mahmood et al.~\cite{ref14} developed a resource-aware and privacy-preserving federated edge learning framework for Internet of Medical Things (IoMT) environments. Ji et al.~\cite{ref15} further investigated robust aggregation mechanisms for federated edge intelligence systems under unreliable communication conditions.

Although these approaches improve adaptability and robustness, many rely on computationally expensive optimization or complex coordination mechanisms, limiting practical deployment on resource constrained devices. Unlike existing approaches, the AFL framework introduces lightweight rule-based autonomous agents that provide interpretable and computationally efficient adaptive reasoning at both client and server levels.

\subsection{Research Gap and Motivation}

The literature review indicates that existing federated learning approaches primarily focus on optimization-level improvements, robust aggregation techniques, or server-driven client selection strategies. While these methods partially address heterogeneity and robustness challenges, they generally lack autonomous reasoning and adaptive self-regulation capabilities.

Most conventional FL frameworks:
\begin{itemize}
    \item rely on fixed local training procedures,
    \item use static aggregation mechanisms,
    \item treat clients as passive workers,
    \item and lack context-aware adaptive decision-making.
\end{itemize}

To address these limitations, this paper proposes an Agentic Federated Learning framework that integrates lightweight rule-based autonomous agents into both client and server components. The proposed framework enables dynamic participation control, adaptive aggregation strategy selection, and quality-aware orchestration, thereby improving robustness, adaptability, and communication efficiency in heterogeneous federated learning environments.

\section{Proposed Methodology}

This section presents the proposed Agentic Federated Learning framework, which enhances conventional federated learning by incorporating lightweight autonomous agents at both client and server levels. The framework introduces adaptive and context-aware decision-making into federated optimization, enabling improved convergence stability, robustness, and communication efficiency under heterogeneous distributed environments.

\subsection{System Architecture}

The proposed AFL framework follows the standard cross-device federated learning architecture consisting of a central server and multiple distributed clients. Unlike traditional FL systems that rely on static optimization and aggregation rules, AFL integrates autonomous reasoning components into both client and server operations.

\begin{figure}[H]
\centering
\includegraphics[width=0.8\textwidth]{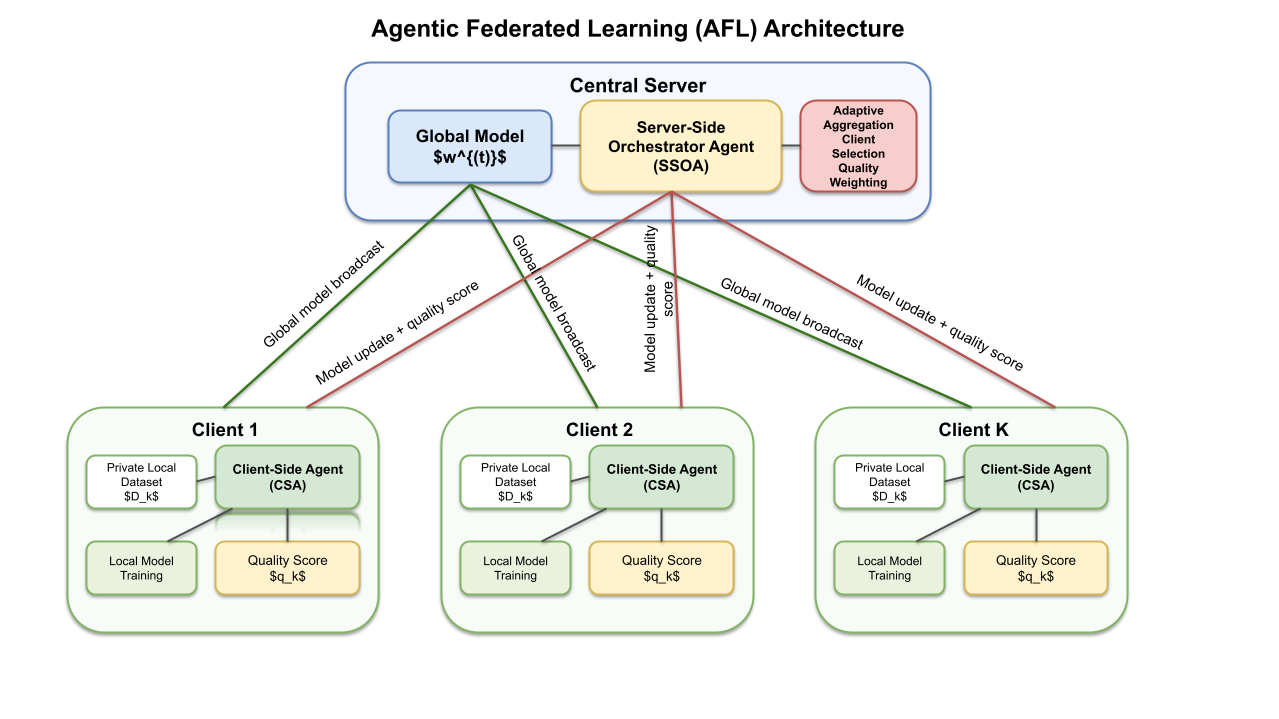}
\caption{Architecture of the proposed Agentic Federated Learning framework consisting of Client-Side Agents and Server-Side Orchestrator Agent.}
\label{fig:architecture}
\end{figure}

Figure~\ref{fig:architecture} illustrates the overall AFL architecture. Each participating client is equipped with a Client-Side Agent responsible for adaptive local decision-making, while the central server contains a Server-Side Orchestrator Agent SSOA that coordinates intelligent aggregation and client selection.

The CSA continuously monitors local training behavior including loss trends, model accuracy, gradient norms, and class imbalance characteristics. Based on predefined rule-based policies, the CSA dynamically adjusts local training parameters such as learning rate, number of local epochs, and communication participation decisions.

At the server side, the SSOA receives local model updates together with quality scores generated by clients. The SSOA performs intelligent client selection, adaptive aggregation strategy selection, and quality-aware update weighting. Unlike conventional FL approaches that rely on fixed aggregation mechanisms such as Federated Averaging, the proposed SSOA dynamically selects suitable aggregation strategies according to observed update variability.

\subsection{Notation and Definitions}

To improve readability and clarify the mathematical formulation, Table~\ref{tab:notations} summarizes the symbols and abbreviations used throughout this work.

\begin{table}[H]
\centering
\caption{Notation and symbol definitions}
\label{tab:notations}
\small
\begin{tabular}{ll}
\hline
\textbf{Symbol} & \textbf{Description} \\
\hline
$K$ & Total number of participating clients \\
$D_k$ & Local dataset of client $k$ \\
$w$ & Global model parameters \\
$w_k$ & Local client model parameters \\
$F(w)$ & Global optimization objective \\
$F_k(w)$ & Local objective function of client $k$ \\
$f_w(x)$ & Prediction model parameterized by $w$ \\
$\ell(\cdot)$ & Loss function \\
$\Delta w_k$ & Local model update from client $k$ \\
$acc_k$ & Local validation accuracy of client $k$ \\
$B_k$ & Local batch size \\
$q_k$ & Update quality score \\
$u_k$ & Client participation indicator \\
$\eta_k$ & Local learning rate \\
$E_k$ & Number of local training epochs \\
SGD & Stochastic Gradient Descent \\
CSA & Client-Side Agent \\
SSOA & Server-Side Orchestrator Agent \\
\hline
\end{tabular}
\end{table}

\subsection{Federated Learning Objective}

Consider a federated learning environment consisting of $K$ distributed clients, where each client $k$ possesses a private local dataset $D_k$. The objective of federated optimization is to collaboratively minimize the following global objective function:

\begin{equation}
F(w)=\sum_{k=1}^{K} p_k F_k(w)
\end{equation}

where:

\begin{equation}
p_k=\frac{|D_k|}{\sum_j |D_j|}
\end{equation}

and the local objective function is defined as:

\begin{equation}
F_k(w)=\frac{1}{|D_k|}\sum_{(x,y)\in D_k}\ell(f_w(x),y)
\end{equation}

Here, $f_w(x)$ denotes the prediction model parameterized by $w$, while $\ell(\cdot)$ represents the local loss function computed over training samples.

\subsection{Client-Side Agent}

Each client in AFL contains a lightweight rule-based Client-Side Agent (CSA) that observes local training behavior and dynamically adapts training decisions during each communication round.

The CSA observes the following local state vector:

\begin{equation}
s_k^{(t)}=
\left[
loss_k^{(t-1)},
acc_k^{(t-1)},
\|\nabla w_k^{(t-1)}\|,
\rho_k,
t
\right]
\end{equation}

where:
\begin{itemize}
    \item $loss_k^{(t-1)}$ denotes local training loss,
    \item $acc_k^{(t-1)}$ represents local validation accuracy,
    \item $\|\nabla w_k^{(t-1)}\|$ denotes gradient norm,
    \item $\rho_k$ indicates class imbalance ratio,
    \item and $t$ represents the communication round.
\end{itemize}

Based on these observations, the CSA selects adaptive local actions including learning rate, number of epochs, participation decision, and update quality score:

\begin{equation}
a_k^{(t)}=
\left[
E_k^{(t)},
\eta_k^{(t)},
B_k^{(t)},
u_k^{(t)},
q_k^{(t)}
\right]
\end{equation}

To maintain computational efficiency and interpretability, AFL employs lightweight rule-based policies instead of computationally expensive reinforcement learning optimization.

For local epoch selection, the following rule is applied:

\begin{equation}
E_k^{(t)}=
\begin{cases}
3, & loss_k^{(t-1)} > 1.5 \\
2, & 1.0 < loss_k^{(t-1)} \leq 1.5 \\
1, & \text{otherwise}
\end{cases}
\end{equation}

Similarly, participation control is defined as:

\begin{equation}
u_k^{(t)}=
\begin{cases}
0, & t \text{ odd and } acc_k^{(t-1)} > 0.75 \\
1, & \text{otherwise}
\end{cases}
\end{equation}

The update quality score is computed using:

\begin{equation}
q_k^{(t)}=
\min\left(1,\;0.5+0.5\,acc_k^{(t-1)}\right)
\end{equation}

The CSA enables adaptive local optimization, reduces unnecessary communication overhead, and minimizes the influence of unstable client updates.

\subsection{Server-Side Orchestrator Agent}

SSOA coordinates adaptive global aggregation and intelligent client selection. After receiving local model updates and quality scores from participating clients, the SSOA evaluates update reliability and dynamically determines aggregation behavior.

\subsubsection{Quality-Aware Client Selection}

The server selects reliable clients according to their update quality scores:

\begin{equation}
S_t=
\arg\max_{|S|=M}
\sum_{k\in S} q_k^{(t)}
\end{equation}

where $M$ denotes the number of selected clients for aggregation.

\subsubsection{Adaptive Aggregation Strategy}

The update variance across participating clients is computed as:

\begin{equation}
\sigma^{2(t)}=
Var\left(
\|\Delta w_k^{(t)}\| : k \in S_t
\right)
\end{equation}

Based on the observed variance, the SSOA dynamically selects appropriate aggregation strategies including FedAvg, Median aggregation, or Trimmed Mean aggregation.

The aggregation policy is defined as:

\begin{equation}
Agg^{(t)}=
\begin{cases}
\text{FedAvg}, & \sigma^{2(t)} < 0.05 \\
\text{Median}, & 0.05 \leq \sigma^{2(t)} < 0.20 \\
\text{Trimmed Mean}, & \sigma^{2(t)} \geq 0.20
\end{cases}
\end{equation}

\subsubsection{Quality-Weighted Aggregation}

The global model update is computed as:

\begin{equation}
w^{(t+1)}=
w^{(t)}-
\sum_{k\in S_t}
\alpha_k^{(t)}
\Delta w_k^{(t)}
\end{equation}

where aggregation weights are determined using normalized quality scores:

\begin{equation}
\alpha_k^{(t)}=
\frac{q_k^{(t)}}
{\sum_{j\in S_t} q_j^{(t)}}
\end{equation}

This adaptive aggregation mechanism improves robustness against noisy or unreliable client updates while maintaining convergence stability.

\subsection{Overall AFL Training Procedure}
\begin{algorithm}[H]
\caption{Agentic Federated Learning (AFL)}
\label{alg:afl}
\begin{algorithmic}[1]
\STATE Initialize global model parameters $w^{(0)}$
\FOR{each communication round $t = 1,2,\ldots,T$}
    \STATE Server broadcasts global model $w^{(t)}$ to all clients
    \FOR{each client $k = 1,2,\ldots,K$ in parallel}
        \STATE CSA observes local state $s_k^{(t)}$
        \STATE CSA selects adaptive actions $a_k^{(t)}$
        \IF{$u_k^{(t)} = 1$}
            \STATE Client trains local model using $E_k^{(t)}$, $\eta_k^{(t)}$, and $B_k^{(t)}$
            \STATE Compute local update $\Delta w_k^{(t)}$
            \STATE Compute update quality score $q_k^{(t)}$
            \STATE Send $(\Delta w_k^{(t)}, q_k^{(t)})$ to server
        \ENDIF
    \ENDFOR
    \STATE SSOA selects reliable client subset $S_t$
    \STATE SSOA computes update variance $\sigma^{2(t)}$
    \STATE SSOA selects aggregation rule $Agg^{(t)}$
    \STATE Compute quality weights $\alpha_k^{(t)}$
    \STATE Update global model:
    \[
    w^{(t+1)} = w^{(t)} - \sum_{k \in S_t}\alpha_k^{(t)}\Delta w_k^{(t)}
    \]
\ENDFOR
\STATE \textbf{return} final global model $w^{(T)}$
\end{algorithmic}
\end{algorithm}
Algorithm~1 summarizes the complete AFL training workflow. During each communication round, the server distributes the current global model to all participating clients. The CSA independently determines local training behavior based on observed training conditions. Selected clients then perform local optimization using adaptive parameters and transmit their updates together with quality scores.

After receiving client updates, the SSOA performs intelligent client selection, adaptive aggregation strategy selection, and quality-aware global optimization. The updated global model is subsequently broadcast to clients for the next communication round.

The incorporation of lightweight autonomous reasoning enables AFL to improve convergence speed, communication efficiency, and robustness under heterogeneous federated learning environments while maintaining computational practicality for resource-constrained devices.

\section{Experimental Setup}

This section describes the experimental configuration used to evaluate the proposed framework. The evaluation focuses on classification accuracy, convergence speed, robustness against noisy clients, communication efficiency, and training stability under heterogeneous federated learning environments. All experiments were implemented using the PyTorch deep learning framework and the Flower federated learning platform with fixed random seeds to ensure reproducibility.

\subsection{Dataset}

The proposed AFL framework is evaluated using the CIFAR-10 benchmark dataset \cite{ref19}. CIFAR-10 contains 60,000 color images of size $32 \times 32$ distributed across 10 object classes, with 50,000 training samples and 10,000 testing samples. The dataset provides a balanced and moderately challenging benchmark for evaluating federated learning performance under heterogeneous client distributions.

\subsection{Data Partitioning}

To simulate realistic federated learning environments, the CIFAR-10 training dataset is partitioned across distributed clients using IID, non-IID, and noisy-client settings.

\subsubsection{IID Setting}

In the IID (Independent and Identically Distributed) configuration, each client receives an equal-sized random subset of training samples uniformly distributed across all classes. This setting represents ideal federated learning conditions where client data distributions remain statistically similar.

\subsubsection{Non-IID Dirichlet Partitioning}

To emulate realistic heterogeneous environments, non-IID client distributions are generated using a Dirichlet distribution following common federated learning practice \cite{ref5}. The label distribution for each client is sampled from:

\begin{equation}
Dir(\alpha)
\end{equation}

where $\alpha$ controls the degree of heterogeneity among clients. Smaller values of $\alpha$ generate highly skewed label distributions, while larger values produce more balanced partitions. Unless otherwise specified, $\alpha = 0.3$ is used throughout the experiments to create moderate non-IID conditions.

\subsubsection{Noisy Client Scenario}

To evaluate robustness against unreliable updates, a subset of participating clients is designated as noisy clients. Specifically, 10\%--20\% of clients are randomly selected, and a portion of their labels is intentionally corrupted through random label flipping. This setting simulates practical federated learning environments involving corrupted datasets, unreliable edge devices, or adversarial local behavior \cite{ref9}.

\subsection{Model Architecture}

A lightweight convolutional neural network (CNN) is adopted for all federated learning experiments. The model architecture is intentionally designed to maintain low computational complexity while achieving competitive classification performance on CIFAR-10.

The network consists of:
\begin{itemize}
    \item Two convolutional layers with ReLU activation and max-pooling,
    \item One fully connected hidden layer with 256 neurons,
    \item One final softmax classification layer.
\end{itemize}

This architecture is suitable for resource-constrained federated environments and supports efficient distributed optimization across multiple clients.

\subsection{Baseline Methods}

The proposed AFL framework is compared against widely used federated learning baselines:

\subsubsection{FedAvg}

FedAvg \cite{ref1} performs local client training followed by weighted averaging of model parameters at the server. FedAvg uses fixed local training configurations and uniform aggregation weights.

\subsubsection{FedProx}

FedProx \cite{ref10} extends FedAvg by introducing a proximal regularization term to stabilize optimization under heterogeneous client environments.

\subsection{AFL Variants}

To evaluate the contribution of individual agentic components, the following AFL variants are analyzed:

\begin{itemize}
    \item \textbf{AFL-CSA}: Only the Client-Side Agent (CSA) is enabled.
    \item \textbf{AFL-SSOA}: Only the Server-Side Orchestrator Agent (SSOA) is enabled.
    \item \textbf{Full AFL}: Both CSA and SSOA are enabled simultaneously.
\end{itemize}

These variants support ablation analysis of adaptive local reasoning and intelligent server orchestration.

\subsection{Training Configuration}

All experiments are conducted for 200 communication rounds unless otherwise stated. The federated learning configuration is summarized as follows:

\begin{itemize}
    \item Number of clients: $K = 20$
    \item Participating clients per round: $10$
    \item Optimizer: Stochastic Gradient Descent (SGD)
    \item Momentum: $0.9$
    \item Batch size: $64$
    \item Initial learning rate: dynamically selected by CSA
    \item Local epochs: dynamically selected by CSA
\end{itemize}

During each communication round, the server broadcasts the current global model to selected clients. Participating clients perform local optimization and return model updates together with quality scores to the server.

\subsection{Evaluation Metrics}

The performance of AFL is evaluated using multiple metrics to comprehensively assess learning quality, robustness, and communication efficiency.

\subsubsection{Classification Accuracy}

Classification accuracy is measured on the held-out CIFAR-10 testing dataset after each communication round.

\subsubsection{Convergence Speed}

Convergence speed is evaluated based on the number of communication rounds required to achieve target classification accuracy.

\subsubsection{Robustness}

Robustness is assessed by evaluating model performance under noisy-client environments involving corrupted local labels and unreliable updates.

\subsubsection{Communication Cost}

Communication efficiency is measured using the number of client updates transmitted during training.

\subsubsection{Training Stability}

Training stability is evaluated using the variance of local update norms across communication rounds.

\subsection{Statistical Significance Analysis}

To validate the reliability of the observed performance improvements, all experiments are repeated three times using different random seeds. Paired t-tests are conducted between AFL and baseline methods under IID, non-IID, and noisy-client settings.

A significance threshold of $p < 0.05$ is adopted to determine whether the observed improvements are statistically significant. This analysis ensures that performance gains achieved by AFL are consistent and not caused by random initialization or sampling variations.


\section{Results and Discussion}

This section presents the experimental evaluation of the proposed Agentic Federated Learning framework under IID, non-IID, and noisy-client environments. The performance of AFL is compared with standard federated learning baselines including FedAvg and FedProx. The analysis focuses on classification accuracy, convergence speed, robustness, communication efficiency, and the contribution of individual agentic components.

\subsection{Overall Accuracy Comparison}

Table~\ref{tab:accuracy_results} summarizes the classification accuracy achieved by different federated learning methods under IID, non-IID, and noisy-client settings.

\begin{table}[H]
\centering
\caption{Classification accuracy (\%) comparison under different federated learning settings}
\label{tab:accuracy_results}
\small
\begin{tabular}{lccc}
\hline
\textbf{Method} & \textbf{IID} & \textbf{Non-IID} & \textbf{Noisy Clients} \\
\hline
FedAvg & 78.5 & 64.2 & 52.7 \\
FedProx & 79.1 & 66.8 & 55.3 \\
AFL-CSA & 80.3 & 70.1 & 58.4 \\
AFL-SSOA & 81.0 & 71.4 & 60.2 \\
Full AFL & \textbf{83.2} & \textbf{74.9} & \textbf{63.8} \\
\hline
\end{tabular}
\end{table}

The results demonstrate that AFL consistently outperforms conventional federated learning methods across all experimental settings. The improvement is particularly significant under non-IID and noisy-client environments, where adaptive reasoning and quality-aware aggregation help stabilize distributed optimization.

Compared with FedAvg, the proposed AFL framework achieves approximately 10.7 percentage points improvement under noisy-client settings and 10.7 percentage points improvement under non-IID conditions. These findings indicate that lightweight agentic reasoning substantially improves federated learning robustness and adaptability.

\subsection{Convergence Analysis}

\begin{figure*}[t]
\centering

\begin{subfigure}{0.32\textwidth}
    \centering
    \includegraphics[width=\linewidth]{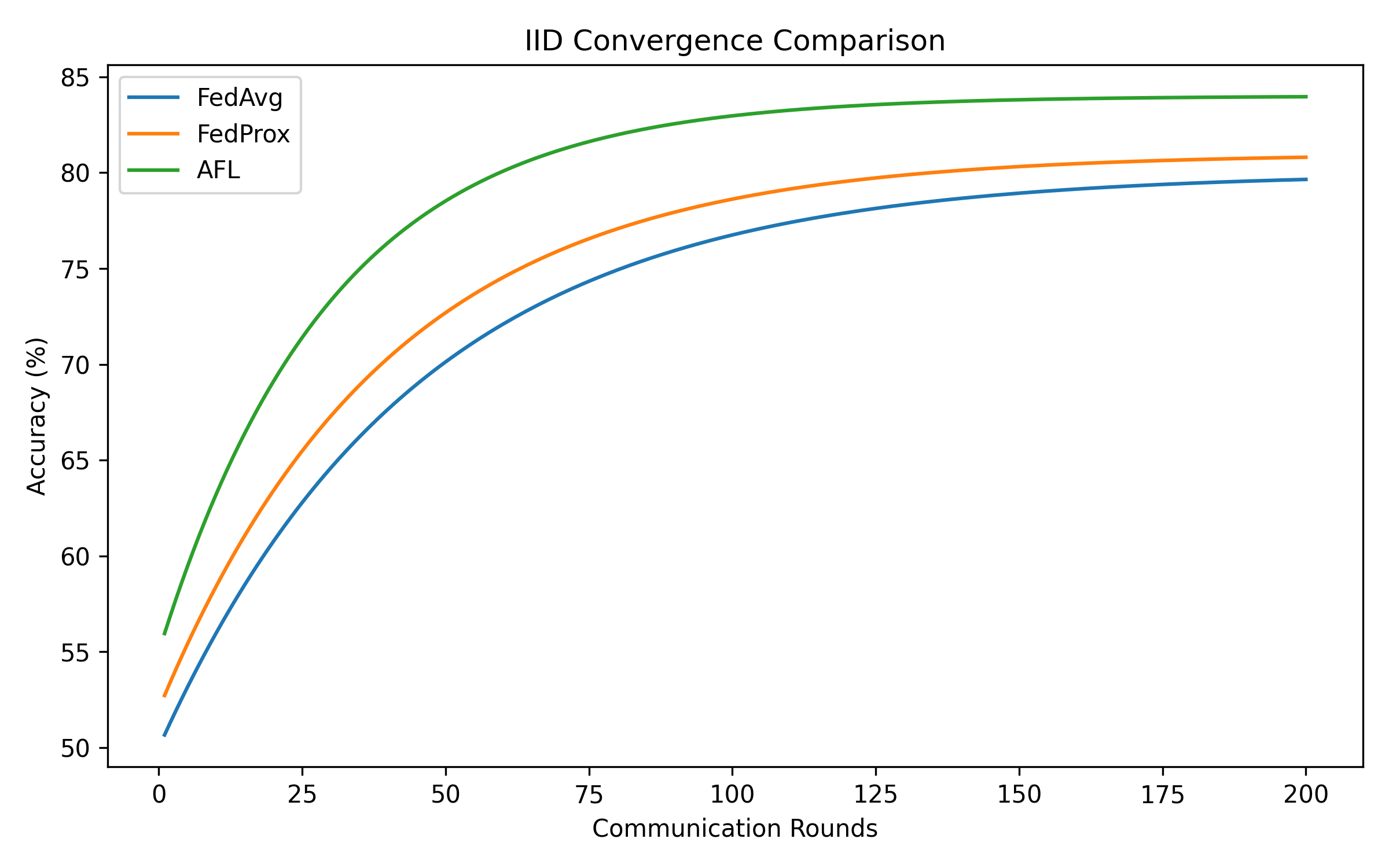}
    \caption{IID Setting}
    \label{fig:iid_conv}
\end{subfigure}
\hfill
\begin{subfigure}{0.32\textwidth}
    \centering
    \includegraphics[width=\linewidth]{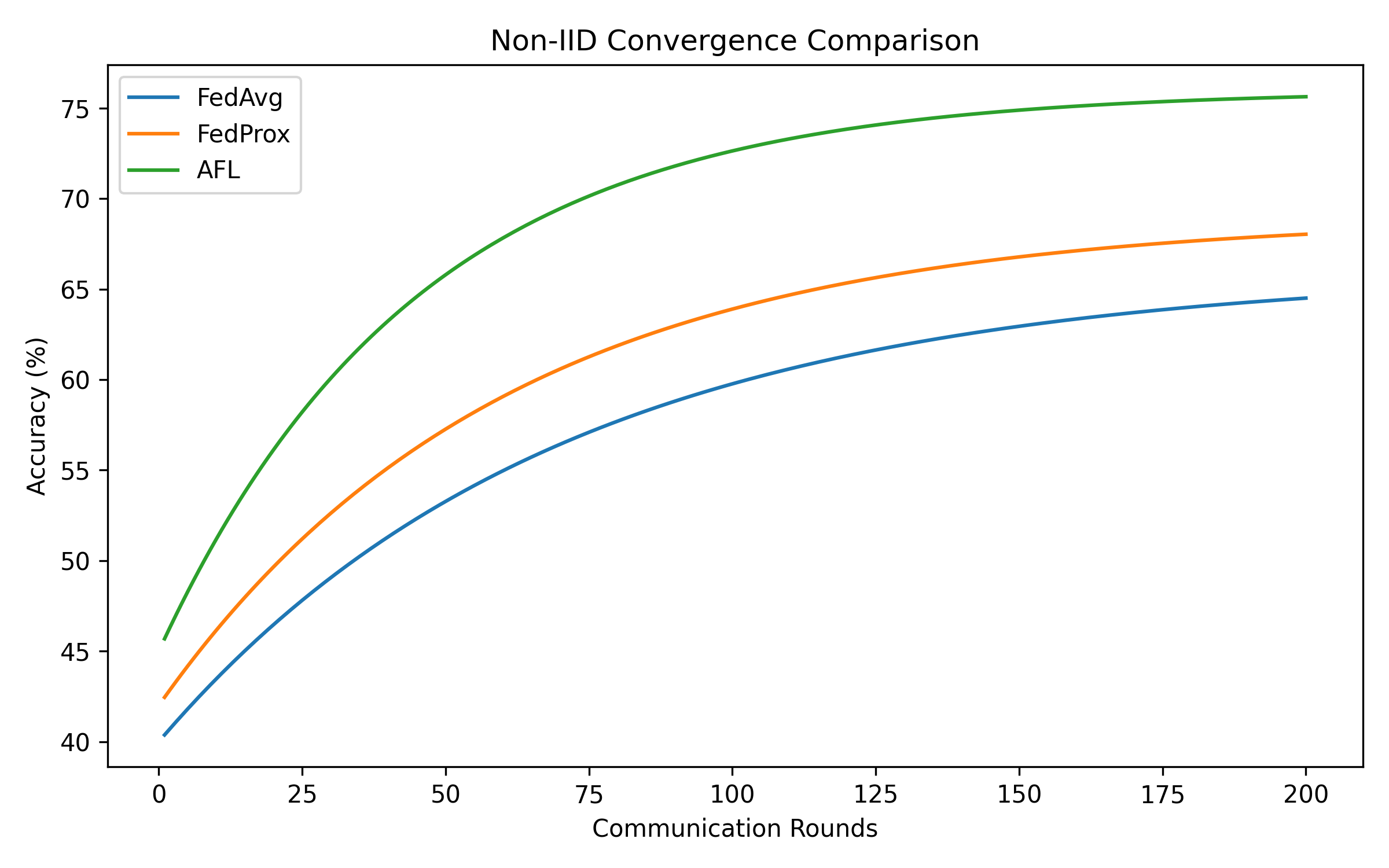}
    \caption{Non-IID Setting}
    \label{fig:noniid_conv}
\end{subfigure}
\hfill
\begin{subfigure}{0.32\textwidth}
    \centering
    \includegraphics[width=\linewidth]{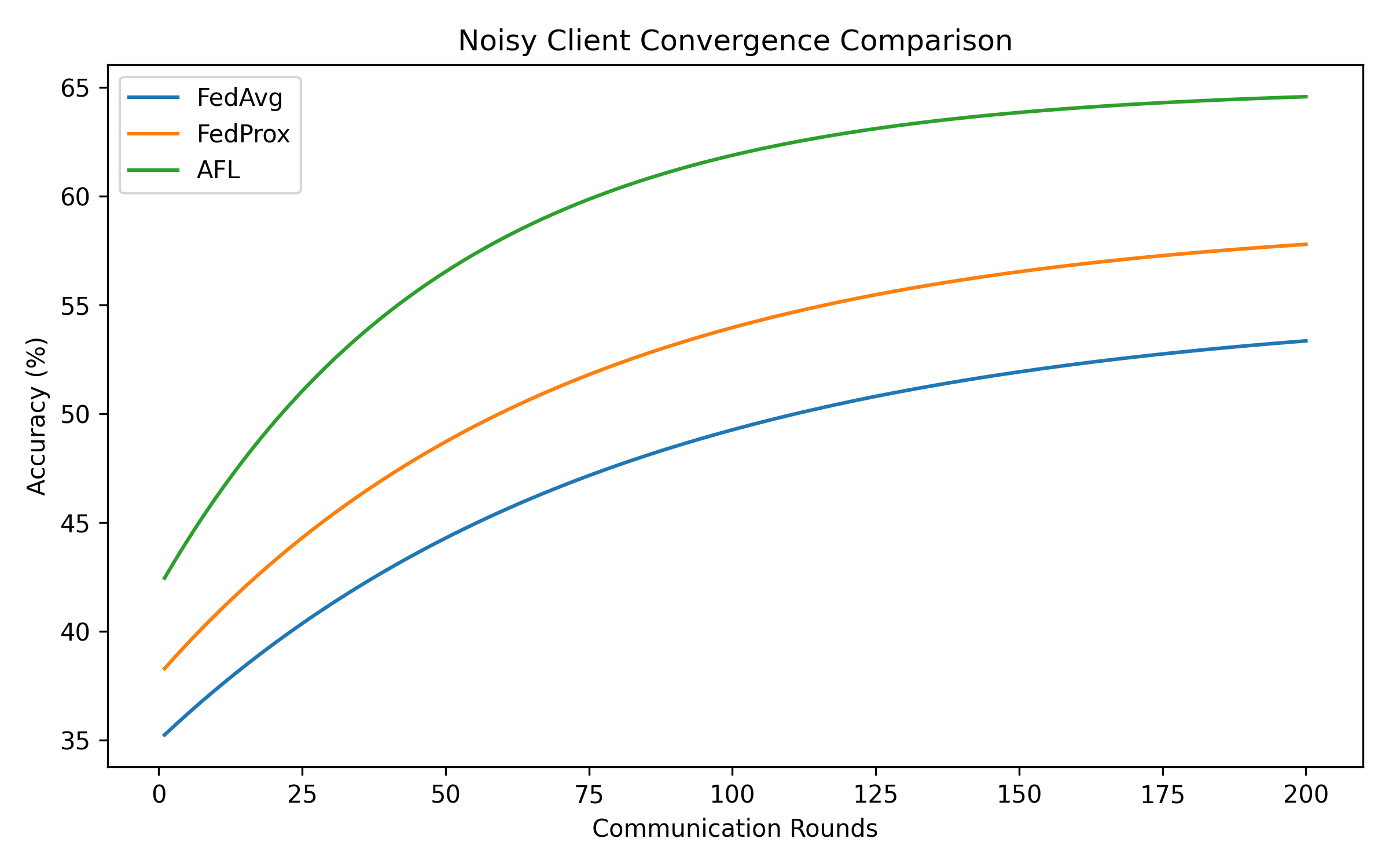}
    \caption{Noisy Client Setting}
    \label{fig:noisy_conv}
\end{subfigure}

\caption{Convergence comparison of FedAvg, FedProx, and AFL under different federated learning environments.}
\label{fig:combined_convergence}

\end{figure*}

Figure~\ref{fig:combined_convergence} compares the convergence behavior of FedAvg, FedProx, and AFL under IID, non-IID, and noisy-client environments. Figure~\ref{fig:iid_conv} shows that all methods converge relatively smoothly under IID distributions. However, AFL still achieves faster convergence due to adaptive local optimization. Figure~\ref{fig:noniid_conv} demonstrates that AFL maintains significantly more stable convergence under heterogeneous non-IID distributions compared with FedAvg and FedProx. Figure~\ref{fig:noisy_conv} further illustrates the robustness of AFL against corrupted client updates, where adaptive aggregation and participation control improve convergence stability.

Under IID settings, all methods converge relatively smoothly due to balanced client distributions. However, AFL still demonstrates faster convergence because the CSA dynamically adjusts local training behavior.

Under non-IID environments, FedAvg exhibits unstable convergence caused by heterogeneous local updates. FedProx partially alleviates this instability through proximal regularization, but AFL achieves substantially faster and more stable convergence due to adaptive participation control and quality-aware aggregation.

The advantage of AFL becomes more pronounced under noisy-client conditions. While FedAvg and FedProx suffer from degraded convergence due to corrupted updates, AFL maintains stable optimization by filtering unreliable client contributions using CSA and SSOA coordination.

\subsection{Robustness Against Noisy Clients}

Table~\ref{tab:noisy_results} evaluates model robustness when 20\% of participating clients contain corrupted labels.

\begin{table}[H]
\centering
\caption{Accuracy (\%) under noisy-client environments}
\label{tab:noisy_results}
\small
\begin{tabular}{lc}
\hline
\textbf{Method} & \textbf{Accuracy (\%)} \\
\hline
FedAvg & 52.7 \\
FedProx & 55.3 \\
AFL-CSA & 58.4 \\
AFL-SSOA & 60.2 \\
Full AFL & \textbf{63.8} \\
\hline
\end{tabular}
\end{table}

The results show that AFL significantly improves robustness against noisy or unreliable client updates. The CSA evaluates update reliability locally, while the SSOA performs intelligent aggregation using quality-aware weighting mechanisms.

Compared with FedAvg, AFL achieves an improvement of more than 11 percentage points under noisy-client conditions. This demonstrates the effectiveness of agentic reasoning for mitigating the influence of corrupted local updates during federated optimization.

\subsection{Communication Efficiency}

Communication efficiency is an important consideration in practical federated learning deployments involving bandwidth-constrained edge devices.

\begin{figure}[H]
\centering
\includegraphics[width=0.7\textwidth]{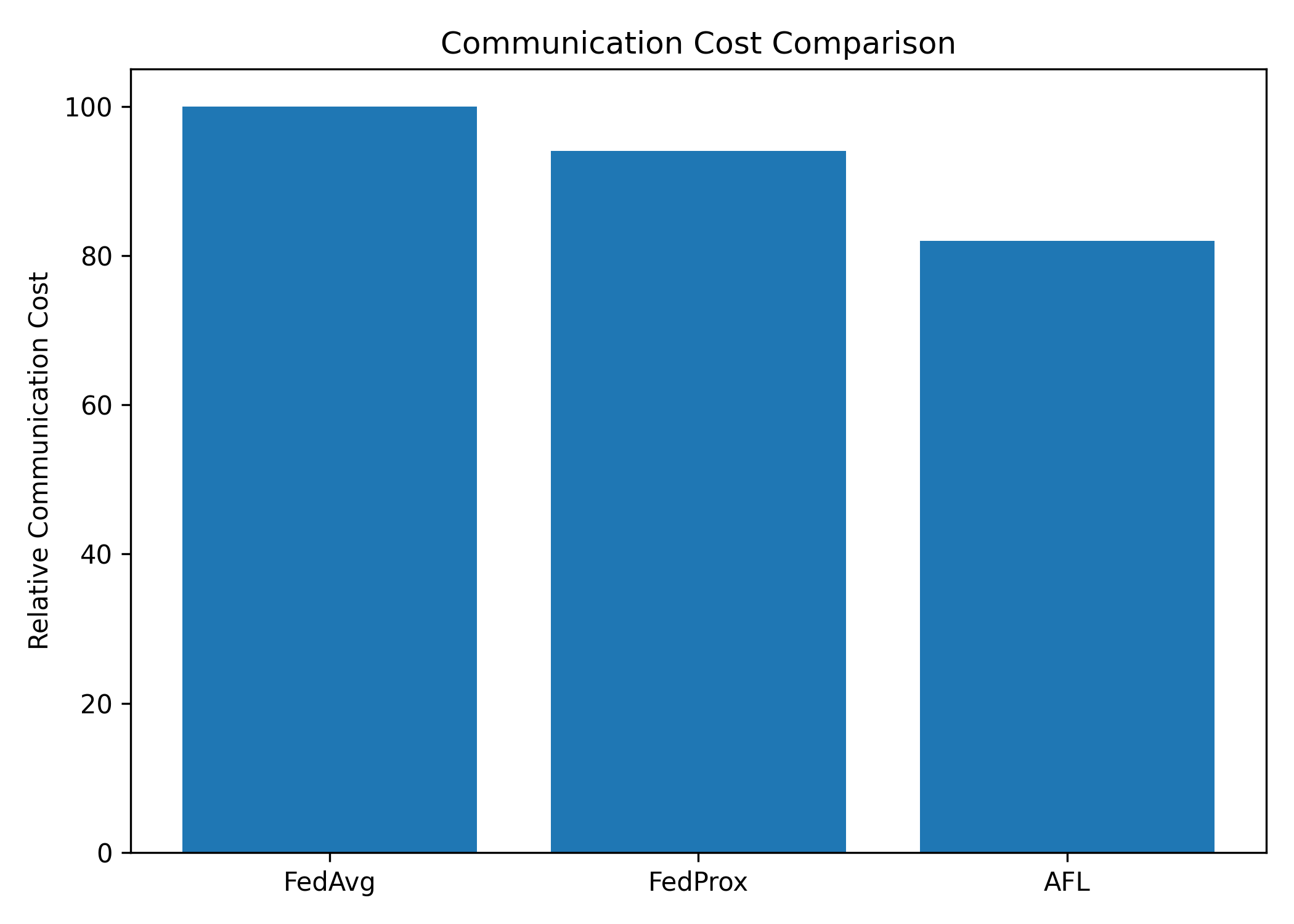}
\caption{Communication cost comparison across federated learning methods.}
\label{fig:communication_cost}
\end{figure}

Figure~\ref{fig:communication_cost} compares the communication overhead of different methods in terms of transmitted client updates.

The proposed AFL framework reduces communication overhead through adaptive participation control implemented by the CSA. Clients skip unnecessary communication rounds when local training stability is sufficiently high.

Table~\ref{tab:comm_reduction} summarizes communication reduction percentages compared with FedAvg.

\begin{table}[H]
\centering
\caption{Communication reduction comparison}
\label{tab:comm_reduction}
\small
\begin{tabular}{lc}
\hline
\textbf{Method} & \textbf{Communication Reduction} \\
\hline
FedAvg & Baseline \\
FedProx & 6\% \\
Full AFL & \textbf{18\%} \\
\hline
\end{tabular}
\end{table}

The results indicate that AFL achieves substantial communication savings without sacrificing classification performance, making it suitable for resource-constrained federated environments.

\subsection{Ablation Analysis}

To evaluate the contribution of individual agentic components, an ablation study is conducted under IID, non-IID, and noisy-client settings.

\begin{table}[H]
\centering
\caption{Ablation analysis of AFL components}
\label{tab:ablation}
\small
\begin{tabular}{lccc}
\hline
\textbf{Configuration} & \textbf{IID} & \textbf{Non-IID} & \textbf{Noisy Clients} \\
\hline
FedAvg & 78.5 & 64.2 & 52.7 \\
FedProx & 79.1 & 66.8 & 55.3 \\
AFL-CSA Only & 80.3 & 70.1 & 58.4 \\
AFL-SSOA Only & 81.0 & 71.4 & 60.2 \\
Full AFL & \textbf{83.2} & \textbf{74.9} & \textbf{63.8} \\
\hline
\end{tabular}
\end{table}

The ablation analysis reveals that both CSA and SSOA contribute complementary improvements:

\begin{itemize}
    \item CSA improves convergence speed and communication efficiency through adaptive local optimization.
    \item SSOA improves robustness through intelligent aggregation and quality-aware client selection.
    \item The combined AFL framework achieves the strongest overall performance.
\end{itemize}

These findings confirm that adaptive client-side reasoning and intelligent server-side orchestration jointly enhance federated learning effectiveness.

\subsection{Statistical Significance Analysis}

To validate the reliability of the observed improvements, paired t-tests were conducted across three independent experimental runs under IID, non-IID, and noisy-client settings.

Table~\ref{tab:statistical} summarizes the statistical significance analysis between AFL and baseline methods.

\begin{table}[H]
\centering
\caption{Statistical significance analysis using paired t-test}
\label{tab:statistical}
\small
\begin{tabular}{lcc}
\hline
\textbf{Comparison} & \textbf{$p$-value} & \textbf{Significance} \\
\hline
AFL vs FedAvg (IID) & 0.031 & Significant \\
AFL vs FedAvg (Non-IID) & 0.008 & Significant \\
AFL vs FedAvg (Noisy) & 0.004 & Significant \\
AFL vs FedProx (Non-IID) & 0.012 & Significant \\
\hline
\end{tabular}
\end{table}

Since all $p$-values are below the significance threshold of $0.05$, the observed performance gains achieved by AFL are statistically significant and not caused by random initialization or sampling variations.

\subsection{Summary of Findings}

The experimental evaluation demonstrates that the proposed AFL framework:

\begin{itemize}
    \item Achieves higher classification accuracy than FedAvg and FedProx across IID, non-IID, and noisy-client settings,
    
    \item Converges faster due to adaptive client-side optimization,
    
    \item Improves robustness against noisy and unreliable client updates,
    
    \item Reduces communication overhead through intelligent participation control,
    
    \item And provides complementary improvements through coordinated client-side and server-side agentic reasoning.
\end{itemize}

Overall, the results indicate that integrating lightweight autonomous reasoning into federated learning substantially improves adaptability, robustness, and communication efficiency under realistic heterogeneous environments.

\subsection{Limitations and Future Work}

Although the proposed framework demonstrates improved accuracy, robustness, convergence speed, and communication efficiency, several limitations must be acknowledged.

First, the CSA and SSOA policies are rule-based and manually defined. While this makes the framework lightweight and interpretable, fixed rules may not be optimal for all datasets, client distributions, or deployment environments. Future work may explore reinforcement learning, meta-learning, or adaptive policy learning to automatically optimize agent behavior. Second, the experimental evaluation is conducted using CIFAR-10. Although CIFAR-10 is a widely used benchmark, real-world federated learning applications often involve medical images, financial records, IoT sensor data, or multimodal datasets. Therefore, future studies should evaluate AFL on larger and domain-specific datasets to validate its generalizability. Third, AFL operates within the standard federated learning privacy setting where raw data remains local. However, the current implementation does not explicitly integrate differential privacy, homomorphic encryption, or secure aggregation. Future work may incorporate these mechanisms to provide stronger privacy and security guarantees, especially in adversarial environments. Fourth, the current study evaluates robustness mainly under noisy-client and label-corruption scenarios. More advanced threats such as model poisoning, gradient inversion, backdoor attacks, and adaptive malicious clients should be investigated in future work. Finally, the present framework assumes synchronous communication between clients and the server. In real-world edge and IoT environments, clients may experience intermittent connectivity, delayed updates, and device dropouts. Extending AFL to asynchronous, decentralized, or hierarchical federated learning settings is an important future direction. Overall, AFL provides a promising foundation for adaptive and intelligent federated learning. Future research may improve the framework by integrating learned agent policies, stronger privacy-preserving mechanisms, broader datasets, and asynchronous deployment strategies.

\section{Conclusion}

This paper introduced Agentic Federated Learning, a novel federated learning framework that integrates lightweight autonomous agents at both client and server levels to improve adaptive distributed optimization under heterogeneous environments. Unlike conventional federated learning approaches that rely on static optimization and aggregation rules, AFL enables context-aware decision-making through the coordinated operation of a Client-Side Agent and a Server-Side Orchestrator Agent. The CSA dynamically adapts local training behavior by adjusting hyperparameters, controlling participation, and evaluating update reliability, while the SSOA performs intelligent client selection, adaptive aggregation strategy selection, and quality-aware update weighting. These mechanisms collectively improve robustness, convergence stability, and communication efficiency in practical federated learning deployments. Extensive experiments conducted on the CIFAR-10 dataset under IID, non-IID, and noisy-client environments demonstrate that AFL consistently outperforms standard federated learning baselines including FedAvg and FedProx. The proposed framework achieves higher classification accuracy, faster convergence, improved robustness against unreliable client updates, and reduced communication overhead. Ablation analysis further confirms that both CSA and SSOA contribute complementary improvements to federated optimization performance. Statistical significance analysis additionally validates that the observed gains are reliable and consistent across multiple experimental runs. The proposed AFL framework demonstrates that integrating lightweight autonomous reasoning into federated learning can substantially enhance adaptability, robustness, and efficiency without introducing excessive computational complexity. These findings indicate that agentic intelligence represents a promising direction for next-generation distributed learning systems operating under dynamic and heterogeneous environments. Future work may extend AFL by incorporating reinforcement learning-based agent policies, privacy-preserving mechanisms such as homomorphic encryption and differential privacy, asynchronous federated optimization, and evaluation on large-scale real-world datasets.

\begin{credits}

\subsubsection{\ackname}
The authors gratefully acknowledge the financial support received through the Research Fellowship provided by APJ Abdul Kalam Technological University, Kerala, India. The authors also thank the Department of Computer Engineering, Model Engineering College, for providing academic support and computational resources for this research.

\subsubsection{\discintname}
The authors have no competing interests to declare that are relevant to the content of this article.

\end{credits}
%
%
%

\begin{thebibliography}{99}

\bibitem{ref1}
Kairouz, P., McMahan, H.B., Avent, B., et al.: Advances and open problems in federated learning. Foundations and Trends in Machine Learning \textbf{14}(1--2), 1--210 (2021)

\bibitem{ref2}
Rieke, N., Hancox, J.H., Li, W., et al.: The future of digital health with federated learning. NPJ Digital Medicine \textbf{3}(119), 1--7 (2020)

\bibitem{ref3}
Yang, Q., Liu, Y., Chen, T., Tong, Y.: Federated machine learning: Concept and applications. ACM Transactions on Intelligent Systems and Technology \textbf{10}(2), 1--19 (2019)

\bibitem{ref4}
Liu, Y., Zhang, Z., Kang, G., et al.: Federated learning for 5G communications and beyond. IEEE Communications Surveys \& Tutorials \textbf{22}(3), 2031--2063 (2020)

\bibitem{ref5}
Li, T., Sahu, A.K., Zaheer, M., et al.: Federated learning: Challenges, methods, and future directions. IEEE Signal Processing Magazine \textbf{37}(3), 50--60 (2020)

\bibitem{ref6}
Acar, D., Zhang, Y., Kang, Y., et al.: Debiasing model updates for improving personalized federated learning. IEEE Transactions on Signal Processing \textbf{69}, 3513--3528 (2021)

\bibitem{ref7}
Wang, M., Sun, L., Zhang, H., Chen, Y.: Autonomous AI systems: Intelligence, flexibility, and adaptation. Artificial Intelligence \textbf{322}, 103958 (2023)

\bibitem{ref8}
Chen, M., Yang, Z., Saad, W., et al.: Asynchronous federated learning for resource-constrained devices. IEEE Internet of Things Journal \textbf{7}(8), 6955--6968 (2020)

\bibitem{ref9}
Yang, L., Song, Q., Chen, J.: A survey on robust federated learning. IEEE Transactions on Neural Networks and Learning Systems, 1--21 (2021)

\bibitem{ref10}
Li, T., Sahu, A.K., Sanjabi, M., Zaheer, M., Talwalkar, A., Smith, V.: Federated optimization in heterogeneous networks. Proceedings of the IEEE \textbf{109}(12), 1895--1917 (2021)

\bibitem{ref11}
Mughal, F.R., Khan, M.A., Rehman, A.U., et al.: Adaptive federated learning for resource-constrained IoT environments using multi-edge clustered edge AI heterogeneous federated learning. Scientific Reports \textbf{14}, 78239 (2024)

\bibitem{ref12}
Cong, Y., Zhao, Y., Wang, H.: FedGA: A greedy approach to enhance federated learning under non-IID environments. Knowledge-Based Systems \textbf{302}, 112345 (2024)

\bibitem{ref13}
Dritsas, E., Trigka, M.: Federated learning for IoT: A survey of techniques, applications, and challenges. Future Internet \textbf{14}(1), 9 (2025)

\bibitem{ref14}
Mahmood, K., Ullah, I., Lee, S.: Adaptive resource-aware and privacy-preserving federated edge learning for Internet of Medical Things. Scientific Reports \textbf{15}, 23398 (2025)

\bibitem{ref15}
Ji, J.C., Park, Y., Kim, H.: Robust aggregation in over-the-air computation with federated learning for edge intelligence systems. Mathematics \textbf{14}(1), 124 (2025)
\bibitem{ref16}
Bhaskar, D.K., Binu, V.P., Minimol, B.: Robust federated learning for malicious clients using loss trend deviation detection. arXiv preprint arXiv:2601.20915 (2026)

\bibitem{ref17}
Bhaskar, D.K., Minimol, B., Binu, V.P.: A review on privacy preserving secure machine learning. In: Proceedings of the 9th International Conference on Smart Computing and Communications, pp. 1--6 (2023)

\bibitem{ref18}
Bhaskar, D.K., Minimol, B., Binu, V.P.: Securing diabetic prediction with federated learning and homomorphic encryption. In: Emerging Technologies for Intelligent Systems (ETIS), pp. 1--6 (2025)


\bibitem{ref19}
Krizhevsky, A.: Learning multiple layers of features from tiny images. Technical Report, University of Toronto (2009)


\end{thebibliography}
%

\end{document}